\documentclass[11pt,a4paper]{article}
\usepackage[T1]{fontenc}
\usepackage[latin9]{inputenc}
\usepackage{amsmath}
\usepackage{amssymb}
\usepackage{esint}

\makeatletter

\pdfoutput=1 

\usepackage{jheppub}

\usepackage{etoolbox}
    \makeatletter
    \patchcmd{\maketitle}{\@fpheader}{}{}{}
    \makeatother

\usepackage{color}

\usepackage{amsfonts}

\usepackage{bbm}

\title{\boldmath Gravitational dual of averaged free CFT's over the Narain lattice}

\author{Alfredo P\'{e}rez,}
\author{Ricardo Troncoso}

\affiliation{Centro de Estudios Cient\'{i}ficos (CECs), Avenida Arturo Prat 514, Valdivia,
Chile.}

\emailAdd{aperez@cecs.cl}
\emailAdd{troncoso@cecs.cl}

\preprint{CECS-PHY-20/01}

\abstract{It has been recently argued that the averaging of free CFT's over
the Narain lattice can be holographically described through a Chern-Simons
theory for $U\left(1\right)^{D}\times U\left(1\right)^{D}$ with a
precise prescription to sum over three-dimensional handlebodies. 

We show that a gravitational dual of these averaged CFT's would be provided
by Einstein gravity on AdS$_{3}$ with $U\left(1\right)^{D-1}\times U\left(1\right)^{D-1}$ gauge
fields, endowed with a precise set of boundary conditions closely
related to the ``soft hairy'' ones. Gravitational excitations then
go along diagonal $SL\left(2,\mathbb{R}\right)$ generators, so
that the asymptotic symmetries are spanned by $U\left(1\right)^{D}\times U\left(1\right)^{D}$
currents. The stress-energy tensor can then be geometrically seen as composite
of these currents through a twisted Sugawara construction. Our
boundary conditions are such that for the reduced phase space, there
is a one-to-one map between the configurations in the gravitational
and the purely abelian theories. The partition function in the bulk
could then also be performed either from a non-abelian Chern-Simons
theory for two copies of $SL\left(2,\mathbb{R}\right)\times U\left(1\right)^{D-1}$
generators, or formally through a path integral along the family of
allowed configurations for the metric. The new boundary conditions
naturally accommodate BTZ black holes, and the microscopic number
 of states then appears to be manifestly positive and suitably accounted for from the
  partition function in the bulk. The inclusion of higher spin currents through
an extended twisted Sugawara construction in the context of higher
spin gravity is also briefly addressed.
}

\makeatother

\begin{document}
\maketitle

\section{Introduction\label{sec:Introduction}}

It has been recently shown that the path integral of Jackiw-Teitelboim
gravity \cite{Teitelboim:1983ux,Jackiw:1984,Teitelboim:1984, Jackiw:1984je} can
be seen as the dual of an average over an ensemble of theories at
the boundary, instead of a single specific one \cite{Saad:2019lba}.
Aiming to lift this result to higher dimensions, in refs. \cite{Maloney:2020nni,Afkhami-Jeddi:2020ezh},
the bulk dual of averaged free two-dimensional CFT's over the Narain
lattice \cite{Narain:1985jj,Narain:1986am} has been identified as
an abelian $U\left(1\right)^{D}\times U\left(1\right)^{D}$ Chern-Simons
theory, given by
\begin{equation}
I=\sum_{I=1}^{D}\int_{Y}\left(A_{I}^{+}dA_{I}^{+}-A_{I}^{-}dA_{I}^{-}\right)\,,\label{eq:Iabelian}
\end{equation}
with a precise prescription to sum over three-dimensional oriented
handlebodies $Y$ with fixed boundary $\partial Y=\Sigma$. The partition
function then fulfills
\begin{equation}
\sum_{Y}Z_{Y}^{U\left(1\right)^{2D}}\left(\tau\right)=\left\langle Z_{\Sigma}\left(m,\tau\right)\right\rangle \,,\label{eq:partitionsfunc}
\end{equation}
where $Z_{\Sigma}\left(m,\tau\right)$ stands for the partition function
of the free CFT$_{2}$ at a point $m$ of the lattice, and $\tau$
collectively denotes the modular parameters of the Riemann surface
$\Sigma$. The bracket $\left\langle \cdots\right\rangle $ stands
for averaging over Narain moduli space, whose measure is determined
by the Zamolodchikov metric, normalized so that the volume of Narain
moduli space is equal to one.

The result can be transparently visualized for a single connected
surface $\Sigma$ of genus $1$. In this case, the partition function
of the free CFT$_{2}$ is given by 
\[
Z_{\Sigma}\left(m,\tau\right)=\frac{\Theta\left(m,\tau\right)}{\left|\eta\left(\tau\right)\right|^{2D}}\,,
\]
where $\eta\left(\tau\right)$ is the Dedekind eta function, and $\Theta\left(m,\tau\right)$
is the ``Siegel-Narain'' theta function. Therefore, by virtue of
the Siegel-Weil formula \cite{siegel1951indefinite,maass1955lectures,weil1964certains,weil1965formule}
the right-hand side of eq. \eqref{eq:partitionsfunc} reads
\begin{equation}
\left\langle Z_{\Sigma}\left(m,\tau\right)\right\rangle =\frac{E_{D/2}\left(\tau\right)}{\text{Im}\left(\tau\right)^{D/2}\left|\eta\left(\tau\right)\right|^{2D}}, \label{Zbdry}
\end{equation}
where $E_{D/2}$ is the real analytic (non-holomorphic) modular invariant
Eisenstein series.

Following \cite{Maloney:2020nni,Afkhami-Jeddi:2020ezh}, the left-hand
side of eq. \eqref{eq:partitionsfunc} can be obtained as follows.
Although the abelian Chern-Simons theory carries no metric in the
bulk, the perturbative contribution to the partition function can
be evaluated on Euclidean (thermal) AdS$_{3}$, whose boundary is
a torus with modular parameter $\tau$. The 1-loop partition function
is then given by that of $D$ left- and right-moving chiral bosons,
determined by the vacuum character of the $U\left(1\right)^{D}\times U\left(1\right)^{D}$
current algebra \cite{Porrati:2019knx}, i.e., 
\[
Z_{\text{EAdS}_{3}}^{U\left(1\right)^{2D}}\left(\tau,\bar{\tau}\right)=\frac{1}{\eta\left(\tau\right)^{D}\eta\left(-\bar{\tau}\right)^{D}}=\chi_{0}\left(\tau\right)\bar{\chi}_{0}\left(\bar{\tau}\right)\,.
\]
For our purposes, it is worth highlighting that this is a direct consequence
of the fact that the asymptotic symmetries of the theory in the bulk
are described by the $U\left(1\right)^{D}\times U\left(1\right)^{D}$
affine algebra.

The full partition function in the bulk, $Z_{\text{bulk}}\left(\tau\right)$,
can then be obtained once the sum over handlebodies is carried out.
This is explicitly performed along the lines of \cite{Maloney:2007ud},
i.e., summing over all modular images in $SL\left(2,\mathbb{Z}\right)$
quotiented by those preserving Euclidean AdS$_{3}$, given by $\Gamma$,
being spanned by $T$-modular transformations. The full partition
function can then be expressed as
\begin{equation}
Z_{\text{bulk}}\left(\tau\right)=\sum_{\gamma\in SL\left(2,\mathbb{Z}\right)/\Gamma}\frac{1}{\left|\eta\left(\gamma\tau\right)\right|^{2D}}=\frac{1}{\text{Im}\left(\tau\right)^{D/2}\left|\eta\left(\tau\right)\right|^{2D}}\sum_{\gamma\in SL\left(2,\mathbb{Z}\right)/\Gamma}\text{Im}\left(\gamma\tau\right)^{D/2},\label{Zbulk}
\end{equation}
which by virtue of the definition of the Eisenstein series, 
\[
E_{s}\left(\tau\right)=\sum_{\gamma\in SL\left(2,\mathbb{Z}\right)/\Gamma}\text{Im}\left(\gamma\tau\right)^{s},
\]
agrees with the average of dual theories at the boundary, $Z_{\text{bulk}}\left(\tau\right)=\left\langle Z_{\Sigma}\left(m,\tau\right)\right\rangle $.

As pointed out in \cite{Maloney:2020nni}, since the theory in the
bulk is abelian, it possesses ``boundary photons'' instead of boundary
gravitons, that are described by chiral and anti-chiral current algebras
at the boundary. Nonetheless, a composite boundary graviton could
always emerge through the Sugawara construction, and it is then natural
to wonder about its gravitational dual in the bulk.

One of the main purposes of our work is showing that a simple setup
to realize the gravitational dual in the bulk can be given by Einstein
gravity on AdS$_{3}$ supplemented by $2\left(D-1\right)$ abelian
gauge fields, endowed with a precise choice of boundary conditions,
being closely related to those in \cite{Afshar:2016wfy,Afshar:2016kjj}.
Our boundary conditions guarantee that the asymptotic symmetries are
spanned by the $U\left(1\right)^{D}\times U\left(1\right)^{D}$ affine
algebra, so that in the reduced phase space there is a one-to-one
correspondence between configurations in the gravitational and abelian
theories.

\section{Gravitational dual and its boundary conditions}

Let us consider Einstein gravity on AdS$_{3}$ with an additional
set of $2\left(D-1\right)$ abelian gauge fields, so that the action
can be written as
\begin{equation}
I\left[g_{\mu\nu},A_{\mu I}^{+},A_{\mu I}^{-}\right]=\frac{1}{16\pi G}\int_{Y}d^{3}x\sqrt{-g}\left(R+2l^{-2}\right)+\sum_{I=2}^{D}\int_{Y}\left(A_{I}^{+}dA_{I}^{+}-A_{I}^{-}dA_{I}^{-}\right)\,.\label{eq:I1}
\end{equation}
It is useful to express the gravitational sector as the difference
of two Chern-Simons actions for $SL\left(2,\mathbb{R}\right)$ with
level $k=l/\left(4G\right)$ \cite{Achucarro:1986vz,Witten:1988hc},
so that the non-abelian gauge fields relate to the dreibein and the
spin connection according to\footnote{The $sl(2,\mathbb{R})$ generators are normalized according to $\left[L_{i},L_{j}\right]=\left(i-j\right)L_{i+j}$,
with $i,j=-1,0,1$, so that the nonvanishing components of the invariant
bilinear form are given by $\left\langle L_{1},L_{-1}\right\rangle =-1$
and $\left\langle L_{0},L_{0}\right\rangle =\frac{1}{2}$.} 
\[
A_{sl(2,\mathbb{R})}^{\pm}=\omega\pm\frac{e}{l}.
\]
The action \eqref{eq:I1} can be written as the difference of two
Chern-Simons actions for $SL(2,\mathbb{R})\times U\left(1\right)^{D-1}$,
up to a boundary term. Following \cite{Coussaert:1995zp} and also
\cite{Henneaux:2013dra,Bunster:2014mua}, it is useful to express
the $SL(2,\mathbb{R})\times U\left(1\right)^{D-1}$ gauge fields $A^{\pm}$
in the asymptotic region as 
\begin{equation}
A^{\pm}=b_{\pm}^{-1}\left(d+a^{\pm}\right)b_{\pm},\label{eq:Agrande}
\end{equation}
where the dependence on the radial coordinate is completely captured
by the group elements $b_{\pm}=b_{\pm}\left(r\right)$. The asymptotic
behavior becomes then fully determined once the excitations of the
auxiliary gauge fields $a^{\pm}$ are specified in the asymptotic
region.

We propose a set of boundary conditions, being such that gravitational
excitations go along the diagonal $sl\left(2,\mathbb{R}\right)$ generators
$L_{0}$, so that 
\begin{equation}
a^{\pm}=\left(\mathcal{J}^{\pm}L_{0}+\sum_{I=2}^{D}\mathcal{J}_{I}^{\pm}Z_{I}\right)dx^{\pm},\label{eq:Achico}
\end{equation}
where $x^{\pm}=\frac{t}{l}\pm\phi$, and $Z_{I}$ stand for the $U\left(1\right)$
generators. Note that the field equations, $F=0$, imply that left
and right excitations are chiral and anti-chiral, respectively.

The asymptotic form of the gauge fields is preserved under gauge transformations
$\delta_{\lambda}a^{\pm}=d\lambda^{\pm}+\left[a^{\pm},\lambda^{\pm}\right]$,
with $\lambda^{\pm}=\eta_{i}^{\pm}L_{i}+\eta_{I}^{\pm}Z_{I}$, provided
that the excitations transform according to 
\begin{align}
\delta\mathcal{J}^{\pm} & =\partial_{\pm}\eta^{\pm}\qquad,\qquad\delta\mathcal{J}_{I}^{\pm}=\partial_{\pm}\eta_{I}^{\pm},\label{eq:transf}
\end{align}
where the parameters $\eta_{I}^{\pm}$ and $\eta^{\pm}:=\eta_{0}^{\pm}$
are (anti-)chiral. Additional gauge transformations spanned by suitable
parameters $\eta_{1}^{\pm}$ and $\eta_{-1}^{\pm}$ can be seen to
be trivial proper gauge transformations that do not contribute to
the canonical generators. The conserved charges can be obtained following
different approaches \cite{Regge:1974zd,Barnich:2001jy}, so that
they are given by $Q=Q^{+}-Q^{-}$, with

\begin{equation}
Q^{\pm}\left[\eta^{\pm},\eta_{I}^{\pm}\right]=\mp\int d\phi\left(\frac{k}{4\pi}\eta^{\pm}\mathcal{J}^{\pm}+\sum_{I=2}^{D}2\eta_{I}^{\pm}\mathcal{J}_{I}^{\pm}\right).\label{eq:Gen}
\end{equation}
The asymptotic symmetry algebra spanned by the generators \eqref{eq:Gen}
can then either be obtained from the computation of their Dirac brackets,
or more quickly, by virtue of $\delta_{\chi}Q\left[\epsilon\right]=\left\{ Q\left[\epsilon\right],Q\left[\chi\right]\right\} $,
from the transformation law of the dynamical fields in \eqref{eq:transf}.
It is readily found to be that of $U\left(1\right)^{D}\times U\left(1\right)^{D}$
currents. Indeed, expanding in Fourier modes according to 
\[
\mathcal{J}^{\pm}=\frac{2}{\sqrt{k}}\sum_{n}J_{n}^{\left(1\right)\pm}e^{\pm in\phi}\qquad,\qquad\mathcal{J}_{I}^{\pm}=\frac{1}{\sqrt{2\pi}}\sum_{n}J_{n}^{\left(I\right)\pm}e^{\pm in\phi},
\]
the nonvanishing brackets read
\begin{equation}
i\left\{ J_{n}^{\left(I\right)\pm},J_{m}^{\left(K\right)\pm}\right\} =\frac{1}{2}n\delta^{IK}\delta_{m+n,0},\label{eq:U1algebra}
\end{equation}
where for the modes, the indices $I,K$ take values on $1,2,\dots,D$.

Our boundary conditions could be regarded as an extension of the ``soft
hairy'' ones in \cite{Afshar:2016wfy,Afshar:2016kjj} for which the
chemical potentials are allowed to depend on the dynamical fields
in a precise way. Different sets of boundary conditions whose chemical
potentials also depend on the dynamical fields have been devised in
order to make contact with two-dimensional integrable systems in refs.
\cite{Perez:2016vqo,Fuentealba:2017omf,Melnikov:2018fhb,Ojeda:2019xih,Grumiller:2019tyl}.

\section{Geometrical emergence of composite Virasoro generators}

Following the lines of \cite{Afshar:2016wfy,Afshar:2016kjj}, a twisted
Sugawara construction naturally emerges from the comparison of the
asymptotic structure of spacetime for our boundary conditions with
that of Brown-Henneaux \cite{Brown:1986nw} endowed with $U\left(1\right)^{D-1}\times U\left(1\right)^{D-1}$
currents. Thus, the asymptotic symmetry algebra of the standard boundary
conditions (with spectral flow), spanned by two copies of the semidirect
sum of Virasoro with the $U\left(1\right)^{D-1}$ affine algebra,
arises from composites of the $U\left(1\right)^{D}\times U\left(1\right)^{D}$ currents.

In order to perform the comparison one has to express both sets of
boundary conditions in terms of the same variables. Note that for
our asymptotic behavior, described through eqs. \eqref{eq:Agrande}
and \eqref{eq:Achico}, the gravitational auxiliary gauge fields $a_{sl\left(2,\mathbb{R}\right)}^{\pm}$
are written in the diagonal gauge, while for the standard Brown-Henneaux
boundary conditions they are expressed in the highest weight gauge
\cite{Coussaert:1995zp}. Therefore, the searched for relationships
can be obtained once the asymptotic form of our connections in \eqref{eq:Agrande}
and \eqref{eq:Achico} are written in the highest weight gauge for
a suitable generic choice of some still unspecified chemical potentials
as in \cite{Henneaux:2013dra,Bunster:2014mua}.

For simplicity, we carry out the comparison just for the ``+'' copy
of the gauge fields. Indeed, one proceeds in the same way for the
other copy, and so we drop the superscript ``+''.

The asymptotic form of the gravitational part of the connection in
the highest weight gauge reads
\[
\hat{A}=\hat{b}^{-1}\left(d+\hat{a}\right)\hat{b},
\]
with $\hat{b}=e^{\rho L_{0}}$, and 
\begin{align*}
\hat{a}_{\phi} & =L_{1}-\frac{1}{2}\left(\mathcal{L}-\frac{4\pi}{k}\sum_{I=2}^{D}\mathcal{J}_{I}^{2}\right)L_{-1},\\
l\hat{a}_{t} & =\mu L_{1}-\mu'L_{0}+\frac{1}{2}\left[\mu''-\left(\mathcal{L}-\frac{4\pi}{k}\sum_{I=2}^{D}\mathcal{J}_{I}^{2}\right)\mu\right]L_{-1},
\end{align*}
where $\mathcal{L}$ and $\mu$ stand for arbitrary functions of $t$
and $\phi$, and prime denotes derivatives with respect to $\phi$.
Note that the spectral flow was already incorporated in the auxiliary
connections, see e.g. \cite{Henneaux:1999ib}.

One can then show that the auxiliary gauge field $\hat{a}$ in the
highest weight gauge and $a$ in the diagonal gauge are related by
a group element $g$ that is permissible in the sense of \cite{Bunster:2014mua},
i.e., it does not interfere with the asymptotic structure. Indeed,
$\hat{a}=g^{-1}\left(d+a\right)g$ with 
\[
g=\exp\left(xL_{1}\right)\exp\left(-\frac{1}{2}\mathcal{J}L_{-1}\right),
\]
where $x=x\left(t,\phi\right)$ fulfills $\mu=l\dot{x}-\mathcal{J}x$,
and $x'-\mathcal{J}x=1$. Consistency with the fact that $\mathcal{J}$
is a chiral mover implies that
\[
\mu'+\mathcal{J}\left(1-\mu\right)=0\,,
\]
which means that the chemical potential $\mu$ at the boundary is
generically given by 
\begin{equation}
\mu\left(t,\phi\right)=1+f\left(t\right)\exp\left[\int_{0}^{\phi}\mathcal{J}\left(\phi'\right)d\phi'\right],\label{eq:mu}
\end{equation}
where $f\left(t\right)$ is an arbitrary function of time.

The gauge fields $a$ and $\hat{a}$ are then mapped into each other
provided that the boundary gravitons are related to the ``boundary
photons'' through a twisted Sugawara construction
\begin{equation}
\mathcal{L}=\frac{4\pi}{k}\sum_{I=2}^{D}\mathcal{J}_{I}^{2}+\frac{1}{2}\mathcal{J}^{2}+\partial_{+}\mathcal{J},\label{eq:Miura}
\end{equation}
which in modes reads

\[
L_{n}=\sum_{I=2}^{D}\left(\sum_{p}J_{n-p}^{\left(I\right)}J_{p}^{\left(I\right)}\right)+\sum_{p}J_{n-p}^{\left(1\right)}J_{p}^{\left(1\right)}+in\sqrt{k}J_{n}^{\left(1\right)}.
\]
Therefore, by virtue of \eqref{eq:U1algebra} the modes $L_{n}$ fulfill
the Virasoro algebra with the Brown-Henneaux central charge
\[
i\left\{ L_{n},L_{m}\right\} =\left(n-m\right)L_{n+m}+\frac{1}{2}kn^{2}\delta_{n+m,0},
\]
together with 
\[
i\left\{ L_{n},J_{m}^{\left(I\right)}\right\} =-mJ_{m+n}^{\left(I\right)},
\]
for $I=2,\dots,D$.

Note that the map between highest weight and diagonal gauge choices
implies that the corresponding parameters $\hat{\eta}$, that preserve
the asymptotic structure in the highest weight gauge, become field
dependent. Indeed, under the action of an asymptotic symmetry spanned
by a purely gravitational $U\left(1\right)$ current, the parameters in the highest weight gauge fulfill 
\begin{equation}
\mathcal{J}\hat{\eta}-\hat{\eta}'=\eta,\label{eq:param}
\end{equation}
so that the transformation law of the boundary gravitons, $\delta\mathcal{L}=2\mathcal{L}\partial_{+}\hat{\eta}+\hat{\eta}\partial_{+}\mathcal{L}-\partial_{+}^{3}\hat{\eta}$,
is recovered from that of the boundary photons $\delta\mathcal{J}=\partial_{+}\eta$
by virtue of the twisted Sugawara relation \eqref{eq:Miura}.

It should be emphasized that, although the currents $L_{n}$ satisfy
the Virasoro algebra, the relationship between the parameters in \eqref{eq:param}
implies that their associated generators $Q\left[\hat{\eta}\right]$
actually span the gravitational $U\left(1\right)$ current algebra,
because 
\[
\delta Q=-\frac{k}{4\pi}\int d\phi\hat{\eta}\delta\mathcal{L}=-\frac{k}{4\pi}\int d\phi\eta\delta\mathcal{J}.
\]

As an ending remark of this section, we point out that since the chemical
potentials $\mu^{\pm}$ are related to the lapse and shift functions
of the boundary metric, they should be single valued. Therefore, eq.
\eqref{eq:mu} implies that if the function $f\left(t\right)$ were
chosen to be non-vanishing, the zero mode of $\mathcal{J}$ should
be excluded. Nonetheless, for the choice $f\left(t\right)=0$ ($\mu=1$),
the zero modes of the $U\left(1\right)$ currents are allowed.

\section{Contact with the abelian theory in the bulk}

The new boundary conditions ensure that the reduced phase space of
the gravitational action \eqref{eq:I1} coincides with that of the
abelian theory in \eqref{eq:Iabelian}. Therefore it is possible to
establish a one-to-one map between configurations in the abelian theory
and those of the gravitational one with our boundary conditions. The
map is precisely given by

\[
A_{U\left(1\right)^{D}}=h^{-1}\left(d+A_{SL(2,\mathbb{R})\times U\left(1\right)^{D-1}}\right)h,
\]
with $h=b^{-1}b_{U\left(1\right)^{D}}$, where $b=b\left(r\right)$
stands for the gauge group element in \eqref{eq:Agrande}, and $b_{U\left(1\right)^{D}}=b_{U\left(1\right)^{D}}\left(r\right)$
corresponds to that of abelian theory. For instance, if the asymptotic
form of the metric were given in normal coordinates, then the gravitational
factor of the group element can be chosen as $b_{\pm}^{sl\left(2,\mathbb{R}\right)}\left(r\right)=\exp\left(\pm\frac{r}{2l}\left(L_{1}-L_{-1}\right)\right)$.

One then concludes that the reduced phase space of the gravitational theory endowed
with our boundary conditions becomes identical to that of the abelian
theory; and therefore, performing the quantization of the reduced phase of both theories
 turns out to be equivalent. 

It is worth pointing out that the partition function in the bulk could then be performed either
from the non-abelian Chern-Simons theory, or formally through a path
integral along a family of allowed configurations for the metric,
i.e. 
\[
\left\langle Z_{\Sigma}\left(m,\tau\right)\right\rangle =\sum_{Y}Z_{Y}^{U\left(1\right)^{2D}}\left(\tau\right)=\sum_{Y}Z_{Y}^{\left(SL(2,\mathbb{R})\times U\left(1\right)^{D-1}\right)^{2}}\left(\tau\right)=\sum_{Y}\int\mathcal{D}g_{\mu\nu}\prod_{I=2}^{D}\mathcal{D}A_{\mu I}e^{-I_{E}},
\]
where $I_{E}$ stands for the Euclidean continuation of the action
\eqref{eq:I1}. Note that the Fadeev-Popov determinant of the gravitational theory should
 be carefully determined so that it precisely captures the relevant contributions from
  gauge fixing that are compatible with our boundary conditions. 
  
It should also be emphasized that the presence of a metric in the bulk
endows $Y$ with a well-defined notion of (Riemannian) manifold with
boundary $\Sigma$.

Our one-to-one map between gravitational and 
abelian configurations has been devised to work in a precise way when $\Sigma$ has the topology of a
torus. It would be interesting to explore how our results could be extended for surfaces $\Sigma$
of higher genus.

\section{Metric formalism, black holes and microscopic counting of states}

Note that the abelian fields in the gravitational theory \eqref{eq:I1}
are described through a Chern-Simons action, so that they do not couple
to the metric; and hence, the spacetime geometry solves the Einstein
equations with negative cosmological constant in vacuum. For our choice
of boundary conditions \eqref{eq:Agrande}, \eqref{eq:Achico}, the
most general solution is given by 

\begin{equation}
ds^{2}=dr^{2}+\frac{l^{2}}{4}\left(\mathcal{J}_{+}^{2}dx^{+2}+\mathcal{J}_{-}^{2}dx^{-2}-2\cosh\left(\frac{2r}{l}\right)\mathcal{J}_{+}\mathcal{J}_{-}dx^{+}dx^{-}\right),\label{eq:metric}
\end{equation}
where $\mathcal{J}^{\pm}=\mathcal{J}^{\pm}\left(x^{\pm}\right)$ are
(anti-)chiral.

Diffeomorphisms that preserve the form of the metric, $\delta_{\xi}g_{\mu\nu}=\mathcal{L}_{\xi}g_{\mu\nu},$
are then spanned by $\xi=\xi^{\mu}\partial_{\mu}$, with $\xi^{r}=0$
and
\[
\xi^{+}=\frac{\eta^{+}}{\mathcal{J}_{+}}\qquad,\qquad\xi^{-}=\frac{\eta^{-}}{\mathcal{J}_{-}}\,,
\]
provided that $\delta\mathcal{J_{\pm}=\partial_{\pm}\eta^{\pm}}$,
with $\partial_{\mp}\eta^{\pm}=0$. The canonical generators associated
to this set of diffeomorphisms fulfill the abelian current algebra.

It is reassuring to verify that the metric and the asymptotic symmetries
agree with those obtained from the Chern-Simons formulation with $g_{\mu\nu}=\frac{l^{2}}{2}\left\langle \left(A_{\mu}^{+}-A_{\mu}^{-}\right)\left(A_{\nu}^{+}-A_{\nu}^{-}\right)\right\rangle $,
and $\eta^{\pm}L_{0}=\xi^{\mu}a_{\mu\,sl(2,\mathbb{R})}^{\pm}$.

Note that generic BTZ black holes \cite{Banados:1992wn,Banados:1992gq}
belong to the reduced phase space, as it can be seen from the zero
modes of \eqref{eq:metric}. Anti-de Sitter spacetime is not in the
(Lorentzian) spectrum, but it is clearly brought back in the Euclidean
continuation. It is worth highlighting that massless and extremal
BTZ black holes become automatically excluded from the allowed configurations,
since the metric degenerates for vanishing $J_{0}^{\left(1\right)+}$
or $J_{0}^{\left(1\right)-}$. Indeed, excluding the latter configurations
turns out to be natural in the Euclidean continuation, since they
possess a different topology that includes an additional boundary.

From the abelian and non-abelian Chern-Simons approaches, excluding
zero modes for the chiral fields appears to be reasonably justified.
Nevertheless, making contact with the bulk metric brings in a puzzling
feature, because zero modes precisely correspond to BTZ black holes
which cannot be so naturally excluded from the spectrum. In fact,
their Bekenstein-Hawking entropy, once expressed in terms of the global
charges reads
\[
S=\frac{A}{4G}=2\pi\sqrt{k}\left(J_{0}^{\left(1\right)+}+J_{0}^{\left(1\right)-}\right),
\]
in agreement with \cite{Afshar:2016wfy,Afshar:2016kjj}, and by virtue
of \eqref{eq:Miura} is equivalent to the Cardy formula supplemented
by the spectral flow, i.e.,
\[
S=\sqrt{2}\pi k\left(\sqrt{\mathcal{L}^{+}-\frac{4\pi}{k}\sum_{I=2}^{D}\left(\mathcal{J}_{I}^{+}\right)^{2}}+\sqrt{\mathcal{L}^{-}-\frac{4\pi}{k}\sum_{I=2}^{D}\left(\mathcal{J}_{I}^{-}\right)^{2}}\right).
\]

Interestingly, the full microscopic number of states appears to be manifestly positive
and suitably accounted for from the partition function in the bulk \eqref{Zbulk}, being equivalent to
averaging over moduli space at the boundary \eqref{Zbdry}. Indeed, as pointed
out in \cite{Afkhami-Jeddi:2020ezh}, the partition function can also be expressed as 

\[
Z_{\text{bulk}}=\sum_{\ell=-\infty}^{\infty}\int_{\left|\ell\right|}^{\infty}d\Delta\rho_{\ell}\left(\Delta\right)\chi_{(\Delta+\ell)/2}\left(\tau\right)\bar{\chi}_{(\Delta-\ell)/2}\left(\bar{\tau}\right),
\]
where 
\[
\rho_{\ell}\left(\Delta\right)=\frac{2\pi^{D}\sigma_{1-D}\left(\ell\right)}{\Gamma\left(D/2\right)^{2}\zeta\left(D\right)}\left(\Delta^{2}-\ell^{2}\right)^{D/2-1}
\]
is the density of non-vacuum primary
 states of spin $\ell=\Delta^{+}-\Delta^{-}$
and scaling dimension $\Delta=\Delta^{+}+\Delta^{-}$ \cite{siegel1951indefinite}, with

\[
\Delta^{\pm}=\frac{k}{2}\left(\mathcal{L}^{\pm}-\frac{4\pi}{k}\sum_{I=2}^{D}\left(\mathcal{J}_{I}^{\pm}\right)^{2}+\frac{1}{2}\right),
\]
and $\sigma_{1-D}\left(\ell\right)$ stands for the sum of $m^{1-D}$ for
all positive integers $m$ dividing $\ell$.

It should be highlighted that  $\rho_{\ell}\left(\Delta\right)$ is manifestly positive and it only corresponds to the density of primaries.
Therefore, once descendants are also included in the counting, the total number of states 
 grows according to the Cardy
formula\footnote{We thank an anonymous referee for pointing out this remark.}. In fact,  the
asymptotic growth of the number of states is clearly dominated by the contribution coming
from the $U(1)$ characters,  while according to $\rho_{\ell}\left(\Delta\right)$, subleading
 contributions necessarily involve left and right movers, so that the full partition function does not holomorphically 
 factorizes. 

\section{Extended Sugawara construction and boundary conditions for higher
spin gravity}

One of the key points to obtain the gravitational dual of the Sugawara
formula was the inclusion of a suitable twisting, precisely as in
the analogue of the Miura map in \eqref{eq:Miura}, so that the stress-energy
tensor fulfills the Virasoro algebra with the Brown-Henneaux central
charge. The precise twisting is geometrically realized in the bulk
for the gravitational theory \eqref{eq:I1} with our boundary conditions.

Due to the presence of the additional $U^{2\left(D-1\right)}$ fields,
it is natural to wonder about a similar construction that allowed
to incorporate higher spin currents. Indeed, for $D>2$, the expected
extended Sugawara construction can be obtained along the lines of
\cite{Grumiller:2016kcp}, so that
\begin{align}
\mathcal{L} & =\frac{k}{4\pi}\left(\frac{2}{3}\mathcal{J}_{\left(3\right)}^{2}+\frac{1}{2}\mathcal{J}^{2}+\partial_{+}\mathcal{J}\right),\nonumber \\
\mathcal{W} & =-\frac{k}{6\pi}\left(-\frac{8}{9}\mathcal{J}_{\left(3\right)}^{3}+2\mathcal{J}^{2}\mathcal{J}_{\left(3\right)}+\mathcal{J}_{\left(3\right)}\partial_{+}\mathcal{J}+3\mathcal{J}\partial_{+}\mathcal{J}_{\left(3\right)}+\partial_{+}^{2}\mathcal{J}_{\left(3\right)}\right),\label{eq:Walgebra}
\end{align}
precisely fulfill the $W_{3}$ algebra with the corresponding central
extension obtained from gravity with a spin three field for the boundary
conditions in \cite{Henneaux:2010xg,Campoleoni:2010zq}. In this case,
the extended Sugawara construction can be seen to emerge from higher
spin gravity with $SL\left(3,\mathbb{R}\right)\times SL\left(3,\mathbb{R}\right)$
gauge group (see e.g. \cite{Blencowe:1988gj,Bergshoeff:1989ns,Vasiliev:1995dn}),
endowed with $U\left(1\right)^{D-2}\times U\left(1\right)^{D-2}$
gauge fields.

Thus, in a generic case, the bulk dual can be given by higher spin
gravity for $SL\left(N,\mathbb{R}\right)\times SL\left(N,\mathbb{R}\right)$
with $U\left(1\right)^{D-N+1}\times U\left(1\right)^{D-N+1}$ gauge fields, with boundary
conditions defined through \eqref{eq:Agrande}, so that the auxiliary
gauge fields in the asymptotic region are given by 
\begin{equation}
a^{\pm}=\left(\mathcal{J}^{\pm}L_{0}+\sum_{s=3}^{N}\mathcal{J}_{\left(s\right)}^{\pm}W_{0}^{\left(s\right)}+\sum_{I=N}^{D}\mathcal{J}_{I}^{\pm}Z_{I}\right)dx^{\pm},\label{eq:Achico-1}
\end{equation}
where $L_{0}$ and $W_{0}^{\left(s\right)}$ are the generators of
the Cartan subalgebra of $SL\left(N,\mathbb{R}\right)$. Therefore,
the asymptotic symmetry algebra can be readily seen to be spanned
by the $U\left(1\right)^{D}\times U\left(1\right)^{D}$ current algebra.

The geometrical realization of the extended Sugawara construction
then naturally emerges from the comparison of our boundary conditions
in the diagonal gauge with those in the highest weight gauge \cite{Henneaux:2010xg,Campoleoni:2010zq,Henneaux:2013dra,Bunster:2014mua}.

\acknowledgments{We thank F\'{a}bio Novaes for useful discussions. This research has been partially supported by Fondecyt grants N$\textsuperscript{\underline{o}}$ 1171162, 1181496, 1181031. The Centro de Estudios Cient\'ificos (CECs) is funded by the Chilean Government through the Centers of Excellence Base Financing Program of Conicyt.}

\bibliographystyle{JHEP}
\bibliography{review}

\end{document}